\documentclass[twocolumn,showpacs,preprintnumbers,pre]{revtex4}
\usepackage{amsmath}
\usepackage{graphicx}
\usepackage{dcolumn}
\usepackage{bm}

\input{tcilatex}

\begin{document}

\section{introduction}

About one and half decades ago, spin models defined on hierarchical lattices
received much attention in the literature \cite{rGK, rSKS, rDEE, rKG, rE,
rDSI, rDIL, rLH1, rLH2}. In general, to construct a hierarchical lattice we
first start with an unit, which may be a bond or a cell, and then proceed a
given type of bond- or cell-decoration iteratively to the infinite limit.
Thus, a hierarchical lattice has fractal structure, and the thermodynamic
limit for a physical system defined on a hierarchical lattice is well
defined. Hierarchical spin models attract researchers' interest mainly due
to two reasons: Firstly theses models are exactly solvable in the context of
the Migdal-Kadanoff renormalization scheme \cite{rMIG, rLPK}. Secondly owing
to the inhomogeneity in the coordination number of lattice sites some
particular properties revealed from the models may provide insights to
inhomogeneous systems such as random magnets, polymers, and percolation
clusters \cite{rBQ}.

$Q$-state Potts model defined on a diamond hierarchical lattice is one
example. Starting with a bond, a diamond hierarchical lattice is obtained by
replacing bonds by diamonds iteratively to the infinite limit. There exists
a remarkable richness of phenomena for the model in the absence of external
fields. In particular, the limiting set of the partition function zeros in
the complex temperature plane, also referred as the Fisher zeros, are
essentially the Julia sets associated with the rational map defined by
renormalization transformation \cite{rDSI, rDIL}, and the Julia set
possesses multifractal structure for $q>0$ \cite{rJKP, rHL1, rHL2}.

The interest about the loci of partition function zeroes has been raised
after the classical works of Yang and Lee on regular lattices \cite{rYL, rLY}%
. After the remarkable Lee-Yang circle theorem, Fisher studied partition
function zeros in the complex temperature plane and showed that the
distribution is an unit circle in the $\sinh \left( 2J/k_{B}T\right) $
complex plane for the two-dimensional zero-field Ising model on simple
square lattice \cite{rF}. Since then, the distributions of Fisher zeros of
the Ising model with isotropic or anisotropic couplings on a variety of
classic planar lattices have been investigated \cite{rSK, rSC, rCHEN, rFEL}.
Recently, Lu and Wu completed the picture by calculating the density of
zeros for two-dimensional Ising model in zero field as well as in a pure
imaginary field $i\pi /2$ on a variety of classic planar lattices \cite{rLW}%
. In principle, by knowing the zeros of the partition function and the
corresponding densities, we may deduce all the thermodynamic characteristics
of a system. For example, the density of the zeros near the critical point
can be used to extract the critical exponents \cite{rSC, rFEL}, and the
logarithmic singularity of the specific heat for the two-dimensional
zero-field Ising model is the result of the linearly vanishing density of
the zeros near the real axis \cite{rF, rLW}.

In this paper we study the distributions and the densities of the Fisher
zeros of the zero-field Ising model on square lattices with diamond-type
bond-decorations, referred as diamond-decorated Ising model (DDIM). The
lattices used in the model are constructed by starting with a simple square
lattice, and then by implemending diamond-type bond-decorations to each bond
iteratively to any desired degree. For DDIM, there exists a well-defined
thermodynamic limit for any finite degree of decorations, and each primary
bond becomes a diamond-hierachical lattice used in the diamond-hierarchical
spin model for the limit of infinite decorations. In our previous work, the
properties of ferromagnetic phase transitions of DDIM have been investigated
extensively for finite as well as infinite decoration-levels \cite{rHLL}.
Here we concentrate on the distribution and the density of the Fisher zeros.

Similar analyses on the distribution of the Fisher zeros have been carried
out for triangular type Ising lattices with cell-decorations \cite{rLHCL}.
These lattices possess the Sierpi{\'{n}}ski gasket as the inherent structure
for a primary triangle in the limit of infinite decoration-level. The
results indicate that the zeros distribution for the infinite decorated
lattices coincides with those for the model defined on the Sierpi{\'{n}}ski
gasket, and the distribution of zeros appears to be an union of infinite
scattered points and a Julia set called the Jordan curve, and the scattered
points are bounded by the Jordan curve. Note that the Jordan curve is a
quasi-one-dimensional circle with the Hausdorff dimension equal to one. It
is also well known that the limiting set of the distribution of the zeros of
the diamond-hierarchical Ising model (DHIM) is a Julia set which owns a
multi-fractal structure \cite{rDSI, rDIL, rJKP, rHL1, rHL2}. The Julia set,
which is a bounded planar distribution with the Hausdorff dimension greater
than one, is quite different from the Jordan curve. Based on these
observations, we may expect that the Julia set occurring in DHIM may also
appear in DDIM, and thence we may understand the formation of the
multi-fractal structure in the Julia set by studying the variation of the
distribution and the density of the zeros in the course of increasing the
decoration-level of DDIM to the infinite limit.

This paper is organized as follows: In Section II, we briefly describe how
to deduce the exact expression of free energy via the construction of the
exact renormalization map of temperature between two successive
decoration-levels and the use of the known results of the Ising model on
simple square lattice. In Section III, we study the distribution of the
Fisher zeros and exhibit the change of the distribution pattern as the
decoration-level increases. In Section IV, we determine the density of the
zeros for the first two decoration-levels by using the results for the case
of simple square lattice, and the densities for higher decoration-levels are
given qualitatively by conjecture. In Section V, we discuss how the Julia
set arise in the limit of infinite decoration-level and characterize its
global mutifractal structure. Finally, Section VI is preserved for summary
and discussion.

\section{free energy}

\label{s2}We construct the exact expression of the free energy of DDIM with
an arbitrary decoration-level $n$ in this section. A simple square lattice
with diamond-type bond-decorations up to the level $n$ is referred as an $n$%
-lattice. Then, simple square lattice itself is $0$-lattice, and its
connecting bonds between any two nearest neighbors are named as $0$-bonds.
We denote the total site-number and bond-number of $0$-lattice as $n_{s}$
and $n_{b}$ with $n_{b}=2n_{s}$. An $1$-bond is formed by replacing a $0$%
-bond by a diamond which consists of four $0$-bonds. Starting with a $0$%
-bond, after the $n$-fold iterative replacements of $0$-bonds with $1$%
-bonds, we obtain an $n$-bond which has site-number $S^{\left( n\right)
}=2(4^{n}+2)/3$ and $0$-bond-number $B^{\left( n\right) }=4^{n}$. An $n$%
-lattice is formed by replacing all $0$-bonds of a $0$-lattice with $n$%
-bonds. For an $n$-lattice, the average site and bond numbers per primary
square of $0$-lattice are $N_{s}^{\left( n\right) }=2S^{\left( n\right) }-3$
and $N_{b}^{\left( n\right) }=2B^{\left( n\right) }$ respectively. The
construction procedure is schematized in Fig. 1.

The general form of the partition function for DDIM defined on an $n$%
-lattice reads 
\begin{equation}
Z^{\left( n\right) }=\sum\limits_{\left\{ \sigma \right\} }\left[
\prod_{\left\langle i,j\right\rangle }\exp \left( \eta \sigma _{i}\sigma
_{j}\right) \right] ,
\end{equation}
where the sum is over all bond-connected pairs $\left\langle
i,j\right\rangle $ of $n$-lattice, and the Ising spin takes two possible
values $\sigma _{i}=\pm 1$. Here we consider uniform ferromagnetic couplings
characterized by the coupling strength $J$, and the dimensionless coupling
parameter $\eta $ is defined as $\eta =J/k_{B}T$.

To calculate the partition function of Eq. (1) for an arbitrary $n$-lattice,
we use the bond-renormalization scheme in evaluating the Boltzmann factors
associated with an $n$-bond. The details of the derivations are given in
Ref. \cite{rHLL}, and we briefly summarize the results in the followings.

The Boltzmann factor associated with the $0$-bond, denoted by $%
B_{\left\langle \mu ,\upsilon \right\rangle }^{\left( 0\right) }$, is given
as $\exp \left( \eta \sigma _{\mu }\sigma _{\nu }\right) $, and it can be
written as 
\begin{equation}
B_{\left\langle \mu ,\upsilon \right\rangle }^{\left( 0\right) }=\cosh
\left( \eta \right) +\sigma _{\mu }\sigma _{\nu }\sinh \left( \eta \right) .
\end{equation}
There are decorated $\left( S^{\left( n\right) }-2\right) $-sites for an $n$%
-bond. The Ising spins defined on the decorated sites couple only to those
belonging to the same $n$-bond, and we refer them as inner spins. Then we
may define the Boltzmann factors associated with an $n$-bond, $%
B_{\left\langle \mu ,\upsilon \right\rangle }^{\left( n\right) }$, as the
result of taking the sum over the inner spins for the product of all
Boltzmann factors associated with the $n$-bond, 
\begin{eqnarray}
B_{\left\langle \mu ,\nu \right\rangle }^{\left( n\right) } &=&\left( \frac{1%
}{2}\right) ^{\left( S^{\left( n\right) }-2\right) }  \notag \\
&&\times \sum_{\sigma _{a},\sigma _{b},...,\sigma _{s}}\exp \left[ \eta
\left( \sigma _{\mu }\sigma _{a}+\sigma _{a}\sigma _{b}+...+\sigma
_{s}\sigma _{\nu }\right) \right] .
\end{eqnarray}
Here the two subscripts, $\mu $ and $\nu $, denote the two primary sites
before decorations, the front factor is added for the normalization of the
sum, and the sum is over the $\left( S^{\left( n\right) }-2\right) $-inner
spins. By substituting the expression of Eq. (2) into each Boltzmann factor
of Eq. (3), we have 
\begin{equation}
B_{\left\langle \mu ,\nu \right\rangle }^{\left( n\right) }=R^{\left(
n\right) }\left( \eta \right) \left[ \cosh \left( \eta ^{\left( n\right)
}\right) +\sigma _{\mu }\sigma _{\nu }\sinh \left( \eta ^{\left( n\right)
}\right) \right] ,
\end{equation}
for $n\geq 1$, where the function $R^{\left( n\right) }\left( \eta \right) $
is given as 
\begin{equation}
R^{\left( n\right) }\left( \eta \right) =\prod_{k=1}^{n}\left[ \exp \left(
\eta ^{\left( k\right) }\right) \right] ^{4^{n-k}},
\end{equation}
and $\eta ^{\left( k\right) }$ is determined by the recursion relation, 
\begin{equation}
\exp \left( \eta ^{\left( k\right) }\right) =\cosh \left( 2\eta ^{\left(
k-1\right) }\right) ,
\end{equation}
with the initial condition $\eta ^{\left( 0\right) }=\eta $ for $1\leq k\leq
n$.

It is well known that the corresponding free energy per bond per $k_{B}T$ of
Eq. (1) for the case of $0$-lattice can be written as 
\begin{eqnarray}
f^{\left( 0\right) } &=&-\frac{1}{4}\ln \sinh \left( 2\eta \right)  \notag \\
&&-\frac{1}{4}\int_{0}^{2\pi }\frac{d\phi }{2\pi }\int_{0}^{2\pi }\frac{%
d\theta }{2\pi }\ln \left[ \sinh \left( 2\eta \right) +\frac{1}{\sinh \left(
2\eta \right) }\right.  \notag \\
&&\left. -\Theta \left( \theta ,\phi \right) \right] ,
\end{eqnarray}
with 
\begin{equation}
\Theta \left( \theta ,\phi \right) =\cos \theta +\cos \phi .
\end{equation}
By observing that up to a factor $R^{\left( n\right) }\left( \eta \right) $
the effective Boltzmann factor of an $n$-bond posesses the same form as that
of a $0$-bond, we can express the free energy density of an $n$-lattice as 
\cite{rHLL} 
\begin{eqnarray}
f^{\left( n\right) } &=&f_{D}^{\left( n\right) }-\frac{1}{4B^{\left(
n\right) }}\ln \sinh \left( 2\eta ^{\left( n\right) }\right)  \notag \\
&&-\frac{1}{4B^{\left( n\right) }}\int_{0}^{2\pi }\frac{d\phi }{2\pi }%
\int_{0}^{2\pi }\frac{d\theta }{2\pi }\ln \left[ \sinh \left( 2\eta ^{\left(
n\right) }\right) \right.  \notag \\
&&\left. +\frac{1}{\sinh \left( 2\eta ^{\left( n\right) }\right) }-\Theta
\left( \theta ,\phi \right) \right] ,
\end{eqnarray}
where $f_{D}^{\left( n\right) }$ is the contribution from the factor $%
R^{\left( n\right) }\left( \eta \right) $, 
\begin{equation}
f_{D}^{\left( n\right) }=-\frac{1}{B^{\left( n\right) }}\ln R^{\left(
n\right) }\left( \eta \right) ,
\end{equation}
which can be expressed as 
\begin{equation}
f_{D}^{\left( n\right) }=-\sum_{k=1}^{n}\frac{1}{4^{k}}\ln \left[ \cosh
\left( 2\eta ^{\left( k-1\right) }\right) \right] ,
\end{equation}
by using the recursion relation of Eq. (6). Note that the $f_{D}^{\left(
n\right) }$ part exists only for $n\geq 1$.

\section{Fisher zeros}

\label{s5} The partition function zeros of a $0$-lattice in the
thermodynamic limit can be obtained by setting the argument of the logarithm
in the free energy density of Eq. (7) equal to zero \cite{rLW}. It is known
that the zeros may lie on the unit circle, $\left| \sinh \left( 2\eta
\right) \right| =1$ \cite{rLW}, or on two circles, $\left| \tanh \left( \eta
\right) \pm 1\right| =\sqrt{2}$ \cite{rF}, depending on the variable used
for the complex temperature plane. In this paper, we study the distribution
and density of the Fisher zeros in the complex $\tanh \left( \eta \right) $\
plane. The basic features appearing in the complex $\sinh \left( 2\eta
\right) $ plane are essentially the same as those we obtain in the complex $%
\tanh \left( \eta \right) $ plane.

By observing the free energy density of Eq. (9), in the thermodynamic limit
we can obtain the zeros distribution of an $n$-lattice from the solutions of
two conditions, 
\begin{equation}
\sinh \left( 2\eta ^{\left( n\right) }\right) +\frac{1}{\sinh \left( 2\eta
^{\left( n\right) }\right) }-\Theta \left( \theta ,\phi \right) =0,
\end{equation}
and 
\begin{equation}
\cosh \left( 2\eta ^{\left( k-1\right) }\right) =0\text{ for }k=1,2,...,n.
\end{equation}
Note that the latter can also be viewed as the condition of the zeros for
the Ising system defined on an $n$-bond, and such a system becomes DHIM in
the limit of infinite $n$.

>From the result of two circles for the partition function zeros of a $0$%
-lattice in the complex $\tanh \left( \eta \right) $ plane, we may conclude
that the solution of Eq. (12) is two circles in the complex $\tanh \left(
\eta ^{\left( n\right) }\right) $ plane, 
\begin{equation}
\left| \tanh \left( \eta ^{(n)}\right) \pm 1\right| =\sqrt{2}.
\end{equation}
The two circles intersect at two points, $i$ and $-i$. As shown in Fig. 2a,
due to the intersections there is a ring contained in the two circles. For
the purpose of identification, we refer the circles as two $n$-cycles and
the ring as $n$-ring. Note that in displaying the distributions of zeros, we
always use bold-faced curves for the right $n$-cycle and its descendants,
and gray curves are for those from the left $n$-cycle. This distribution
also appears to have the symmetry of inversion about the center $\tanh
\left( \eta ^{(n)}\right) =0$.

For the purpose of comparing the zeros distributions among different $n$%
-lattices, we have to bring the zeros to the complex plane of an unique
variable chosen to be $\tanh \left( \eta \right) $. To achieve this, we
notice that the recursion relation of Eq. (6) can be rewritten as 
\begin{equation}
\tanh \left( \eta ^{\left( n\right) }\right) =\frac{2\left[ \tanh \left(
\eta ^{\left( n-1\right) }\right) \right] ^{2}}{1+\left[ \tanh \left( \eta
^{\left( n-1\right) }\right) \right] ^{4}},
\end{equation}
which has the inverse map given as 
\begin{equation}
\tanh \left( \eta ^{\left( n-1\right) }\right) =\pm \left( \frac{1\pm \sqrt{%
1-\left[ \tanh \left( \eta ^{\left( n\right) }\right) \right] ^{2}}}{\tanh
\left( \eta ^{\left( n\right) }\right) }\right) ^{1/2}.
\end{equation}
Then, starting with the two $n$-cycles in the complex $\tanh \left( \eta
^{\left( n\right) }\right) $ plane, we can obtain the corresponding zeros
distribution in the complex $\tanh \left( \eta \right) $ plane by performing
the $n$-fold backward iterations provided by Eq. (16).

After the first backward iteration, we show the resultant distribution in
the complex $\tanh \eta ^{\left( n-1\right) }$ plane in Fig. 3a where the
points indicated by crosses are the preimages of the map of Eq. (15) for the
centers of two $n$-cycles, $1$ and $-1$. The results indicate that each of
the $n$-cycles shown in Fig. 2a splits to two closed curves referred as $%
\left( n-1\right) $-cycles. There are $8$ intersection points between the
descendants of the right $n$-cycle and those from the left $n$-cycle, and
the loci of the intersections are determined by the inverse maps of the
points, $i$ and $-i$, which are the intersections of two $n$-cycles. There
are four rings, referred as $\left( n-1\right) $-rings, caused by the
intersections. Note that the distribution shown in Fig. 3a are just the
zeros distribution in the complex $\tanh \left( \eta \right) $ plane subject
to the condition of Eq. (12) for an $1$-lattice.

Proceeding with the inverse map given by Eq. (16) from the complex variable $%
\tanh \left( \eta ^{\left( n-1\right) }\right) $ to $\tanh \left( \eta
^{\left( n-2\right) }\right) $ for the four $\left( n-1\right) $-cycles, we
obtain $12$ $\left( n-2\right) $-cycles as shown in Fig. 4a. The $\left(
n-2\right) $-cycles contain $16$ $\left( n-2\right) $-rings caused by the $%
32 $ intersections among the cycles. The intersections are again the
preimages of the map of Eq.(15) for the loci of the intersections among the $%
\left( n-1\right) $-cycles.

Continuing with this procedure, we show the distribution of zeros in Fig. 5
for $n=4$ and Fig. 6 for $n=8$. In general, the zero distribution for a $n$%
-lattice in the complex $\tanh \eta $ plane, subject to the condition of Eq.
(12), is the union of $\left[ 2+2\left( 4^{n}-1\right) /3\right] $ $0$%
-cycles which have $4^{n+1}/2$ intersections, and these intersections yield $%
4^{n}$ $0$-rings contained in the $0$-cycles. The $0$-cycles can be divided
into two classes: one consists of the the descendants of the left $n$-cycle
with $\left[ 2+2\left( 4^{n-1}-1\right) /3\right] $ members, and the other
has $4^{n}/2$ members which are the descendants of the right $n$-cycle. The
intersections only occur between two $0$-cycles belonging to different
class. The distribution always maintain the inversion symmetry about the
center $\tanh \eta =0$.

For the condition of Eq. (13), it can be rewritten in terms of the $\tanh
\eta ^{(n)}$ variable as 
\begin{equation}
\tanh \eta ^{(k-1)}=\pm i\text{ for }k=1,2,...,n.
\end{equation}
Then, for the case of $n=1$ the two zeros, $i$ and $-i$, in the complex $%
\tanh \eta $ plane are just the intersection points of two $n$-cycles in the
complex $\tanh \left( \eta ^{\left( 1\right) }\right) $ plane. Proceeding to
the case of $n=2$, we obtain, besides of the previous two zeros, $8$ more
points in the complex $\tanh \left( \eta \right) $ plane from $k=2$ in Eq.
(17). These additional points are the preimages of the renormalization map
for the original two points, and they are the intersection points among four 
$\left( n-1\right) $-cycles in the complex $\tanh \left( \eta ^{\left(
n-1\right) }\right) $ plane. By induction, we may conclude that the zeros in
the complex $\tanh \left( \eta \right) $ plane obtained from the condition
of Eq. (17) are $2\left( 4^{n}-1\right) /3$ scattered points which are the
union of the intersection points among $k$-cycles in the complex $\tanh
\left( \eta ^{\left( k\right) }\right) $ plane for $1\leq k\leq n$. in the
complex $\tanh \eta $ plane for an $n$-lattice.

Thus, we can describe the distribution pattern of the Fisher zeros of DDIM
with the decoration-level $n$ in the complex $\tanh \eta $ plane as follows.
There are $2\left( 4^{n}-1\right) /3$ scattered points given by Eq. (17). In
addition, there are $\left[ 2+2\left( 4^{n}-1\right) /3\right] $ $0$-cycles
with $2\cdot 4^{n}$ intersections obtained from Eq. (12).

Among all the zeros we obtained in the above, as a consequence of the
Lee-Yang theorem \cite{rYL, rLY}, the bulk critical points correspond to the
zeros falling on the physical region. The physical region of the variable $%
\tanh \left( \eta \right) $ is $0\leq \tanh \left( \eta \right) <1$ for
ferromagnetic couplings $\eta \geq 0$. This implies that the variable $\tanh
\left( \eta ^{\left( n\right) }\right) $ also takes the range $0\leq \tanh
\left( \eta ^{\left( n\right) }\right) <1$ as the physical region for any $n$
value.

>From the solutions of the conditions of Eqs. (12) and (13), we know that
there is only one zero in the physical region. This zero belongs to the
right $n$-cycle and locates at 
\begin{equation}
\tanh \left( \eta ^{\left( n\right) }\right) \overset{c}{=}h^{\left(
0\right) },
\end{equation}
with $h^{\left( 0\right) }=\sqrt{2}-1$ for arbitrary decoration-level $n$.
Here, for convenience, we use the notation, $\overset{c}{=}$, to denote the
equality established only at the phase transition point. Note that the $%
h^{\left( 0\right) }$ value is just the reduced critical temperature, $\tanh
\left( J/k_{B}T_{c}\right) $, of the ferromagnetic phase transition for the
square Ising model, i.e. $n=0$.

To find the locus of the zero specified by Eq. (18) in the complex $\tanh
\eta $ plane, we can continuously use the inverse map of Eq. (16) to obtain
the equivalent expression of Eq. (18) as 
\begin{equation}
\tanh \left( \eta ^{\left( n-k\right) }\right) \overset{c}{=}h^{\left(
k\right) },
\end{equation}
with 
\begin{equation}
h^{\left( k\right) }=\left( \frac{1-\sqrt{1-\left( h^{\left( k-1\right)
}\right) ^{2}}}{h^{\left( k-1\right) }}\right) ^{1/2},
\end{equation}
for $1\leq k\leq n$. Note that in obtaining Eq. (20) for the critical value
of $\tanh \eta ^{\left( n-k\right) }$ we have used the constraint $0\leq
\tanh \eta ^{\left( n-k\right) }<1$.

Thus, the zero, $h^{\left( 0\right) }$, in the complex $\tanh \eta ^{\left(
n\right) }$ plane corresponds to the zero, $h^{\left( n\right) }$, in the
complex $\tanh \eta $ plane, and the $h^{\left( n\right) }$ value is just
the reduced critical temperature $\tanh \left( J/k_{B}T_{c}\right) $ of the
ferromagnetic phase transition for DDIM with the decoration-level $n$. The
sequence of $h^{\left( n\right) }$ decreases as $n$ increases, and the $%
h^{\left( n\right) }$ value in the limit of infinite $n$ is given by the
asymptotic value of the sequence of $h^{\left( n\right) }$. For the
recursion relation of Eq. (15), there are three fixed points, one repellor
locating at $0.5437...$, and two attracors at $0$ and $1$. Since the $%
h^{\left( n\right) }$ value is obtained from $h^{\left( 0\right) }$ via the $%
n$-fold backward iterations given by Eq. (20) and the attractors (repellors)
of the map become the repellors (attractors) of the inverse map, we may
conclude that the $h^{\left( n\right) }$ value is the locus of the repellor
of Eq. (15), $h^{\left( n\right) }=0.5437...$, for the case of infinite $n$.

\section{density of zeros}

The density of two $n$-cycles in the complex $\tanh \eta ^{\left( n\right) }$
plane has been determined by Lu and Wu \cite{rLW}. Based on this result, we
determine the density of the zeros of DDIM in this section by performing
proper transformations. First, we describe the result of Lu and Wu briefly
in the following.

The two cycles of Eq. (14) are written as 
\begin{equation}
\tanh \eta ^{\left( n\right) }\pm 1=r\left( \alpha ^{\left( n\right)
}\right) \exp \left( i\alpha ^{\left( n\right) }\right) ,
\end{equation}
where $r\left( \alpha ^{\left( n\right) }\right) =\sqrt{2}$ is the radial
distance from the center coordinate. By considering an $M\times 2N$
simple-quartic lattice with Brascamp-Kunz boundary condition, we introduce
the zero density, $g_{\pm }\left( \alpha ^{\left( n\right) }\right) $, which
satisfies the normalization condition 
\begin{equation}
\int_{0}^{2\pi }g_{\pm }\left( \alpha ^{\left( n\right) }\right) d\alpha
^{\left( n\right) }=\frac{1}{2},
\end{equation}
such that the number of zeros in the interval $\left[ \alpha ^{\left(
n\right) },\alpha ^{\left( n\right) }+d\alpha ^{\left( n\right) }\right] $
is $2MNg_{\pm }\left( \alpha ^{\left( n\right) }\right) d\alpha ^{\left(
n\right) }$ for the left ($+$) and right ($-$) $n$-cycle respectively. The
zero density is given as 
\begin{eqnarray}
g_{+}\left( \alpha ^{\left( n\right) }\right) &=&g_{-}\left( \pi -\alpha
^{\left( n\right) }\right) =\left( \frac{x}{\pi ^{2}}\right) \left| \frac{1-%
\sqrt{2}\cos \alpha ^{\left( n\right) }}{3-2\sqrt{2}\cos \alpha ^{\left(
n\right) }}\right|  \notag \\
&&\times K\left( x\right) ,
\end{eqnarray}
where 
\begin{equation}
x=\frac{2\left| \sin \alpha ^{\left( n\right) }\right| \left( \sqrt{2}-\cos
\alpha ^{\left( n\right) }\right) }{3-2\sqrt{2}\cos \alpha ^{\left( n\right)
}},
\end{equation}
and $K\left( x\right) $ is the complete elliptic integral of the first kind, 
\begin{equation}
K(x)=\int_{0}^{\pi /2}dt\frac{1}{\sqrt{1-x^{2}\sin ^{2}t}}.
\end{equation}
The density of $g_{+}\left( \alpha ^{\left( n\right) }\right) $, which has
the symmetry $g_{+}\left( \alpha ^{\left( n\right) }\right) =g_{+}\left(
2\pi -\alpha ^{\left( n\right) }\right) $, is plotted in Fig. 2b for the
range $0\leq \alpha ^{\left( n\right) }\leq \pi $. Here we also specify the
radial distance of a zero, $r\left( \alpha ^{\left( n\right) }\right) $, in
the right vertical scale. For small $\alpha ^{\left( n\right) }$, the result
of Eq. (23) has the linear behavior as 
\begin{equation}
g_{\pm }\left( \alpha ^{\left( n\right) }\right) =\left( \frac{3\pm 2\sqrt{2}%
}{\pi }\right) \left| \alpha ^{\left( n\right) }\right| .
\end{equation}
Note that the the density $g_{+}\left( \alpha ^{\left( n\right) }\right) $
of the zeros near the the point $\alpha ^{\left( n\right) }=0$ is the same
as the density $g_{-}\left( \alpha ^{\left( n\right) }\right) $ near the
point $\alpha ^{\left( n\right) }=\pi $ which is the ferromagnetic phase
transition point of the bulk system, and this linearly vanishing density of
the zeros near the bulk transition point leads to the logarithmic
singularity of the specifi heat. \ 

To find the corresponding density in the complex $\tanh \left( \eta ^{\left(
n-1\right) }\right) $ plane, we first write the $\left( n-1\right) $-cycles
as 
\begin{equation}
\tanh \eta ^{\left( n-1\right) }=z_{0}+r\left( \alpha ^{\left( n-1\right)
}\right) \exp (i\alpha ^{\left( n-1\right) }).
\end{equation}
Here the center coordinates, $z_{0}$, are choosen to be the preimages of the
map of Eq. (15) for the center coordinates of $n$-cycles, $\tanh \eta
^{\left( n\right) }=1$ and $-1$, and this leads to $z_{0}=1$, $-1$, $i$, and 
$-i$ for the center coordinates of $4$ $\left( n-1\right) $-cycles. The zero
density of the $\left( n-1\right) $-cycle, specified by the center
coordinate $z_{0}$, is denoted as $g_{z_{0}}\left( \alpha ^{\left(
n-1\right) }\right) $, and we have the relation, 
\begin{eqnarray}
g_{1}\left( \alpha ^{\left( n-1\right) }\right) &=&g_{i}\left( \frac{\pi }{2}%
+\alpha ^{\left( n-1\right) }\right) =g_{-1}\left( \pi +\alpha ^{\left(
n-1\right) }\right)  \notag \\
&=&g_{-i}\left( \alpha ^{\left( n-1\right) }-\frac{\pi }{2}\right) ,
\end{eqnarray}
for the distribution shown in Fig. 3a. Thence, we determine the density $%
g_{1}\left( \alpha ^{\left( n-1\right) }\right) $, and the densities of
other $\left( n-1\right) $-cycles are followed from $g_{1}\left( \alpha
^{\left( n-1\right) }\right) $ according to the above relation.

Because that the inverse map given by Eq. (16) is one to four and the $%
\left( n-1\right) $-cycle of $z_{0}=1$ is a descendant of the left $n$%
-cycle, we can express the density $g_{1}\left( \alpha ^{\left( n-1\right)
}\right) $ as 
\begin{equation}
g_{1}\left( \alpha ^{\left( n-1\right) }\right) =\frac{g_{+}\left( \alpha
^{\left( n\right) }\right) }{4}\left\vert \frac{d\alpha ^{\left( n\right) }}{%
d\alpha ^{\left( n-1\right) }}\right\vert .
\end{equation}
To determine the transformation Jacobian, $\left\vert d\alpha ^{\left(
n\right) }/d\alpha ^{\left( n-1\right) }\right\vert $, we first notice that
the $\left( n-1\right) $-cycles of $z_{0}=\pm 1$ are the solutions of the
equation, 
\begin{eqnarray}
&&\left\vert \tanh \eta ^{\left( n-1\right) }\right\vert ^{4}-\left[ \left(
\tanh \eta ^{\left( n-1\right) }\right) ^{2}+\left( \tanh \eta ^{\left(
n-1\right) \ast }\right) ^{2}\right]  \notag \\
&&-2\sqrt{2}\left\vert \tanh \eta ^{\left( n-1\right) }\right\vert ^{2}+1 
\notag \\
&=&0,
\end{eqnarray}
where $\tanh \eta ^{\left( n-1\right) \ast }$ is the complex conjugate of $%
\tanh \eta ^{\left( n-1\right) }$. This result is obtained by substituting
Eq. (15) into Eq. (14) for the left $n$-cycle. By further substituting Eq.
(27) with $z_{0}=1$ into Eq. (30), we obtain 
\begin{eqnarray}
&&r^{4}+r^{3}\left[ 4\cos \left( \alpha ^{\left( n-1\right) }\right) \right]
+r^{2}\left( 4-2\sqrt{2}\right)  \notag \\
&&-r\left[ 4\sqrt{2}\cos \left( \alpha ^{\left( n-1\right) }\right) \right]
-2\sqrt{2}  \notag \\
&=&0.
\end{eqnarray}
This equation can be solved numerically to obtain $r\left( \alpha ^{\left(
n-1\right) }\right) $ and $dr/d\alpha ^{\left( n-1\right) }$. Moreover, the
relation between $\alpha ^{\left( n\right) }$ and $\alpha ^{\left(
n-1\right) }$ has been specified by the map of Eq. (15). We substitute Eqs.
(21) and (27) into Eq. (15) to obtain 
\begin{equation}
1+\sqrt{2}\exp \left( i\alpha ^{\left( n\right) }\right) =\frac{2\left[
1+r\left( \alpha ^{\left( n-1\right) }\right) \exp (i\alpha ^{\left(
n-1\right) })\right] ^{2}}{1+\left[ 1+r\left( \alpha ^{\left( n-1\right)
}\right) \exp (i\alpha ^{\left( n-1\right) })\right] ^{4}}.
\end{equation}
By differentiating this equation with respect to $\alpha ^{\left( n-1\right)
}$ and by using the known values of $r\left( \alpha ^{\left( n-1\right)
}\right) $ and $dr/d\alpha ^{\left( n-1\right) }$, we can determine the
derivative, $d\alpha ^{\left( n\right) }/d\alpha ^{\left( n-1\right) }$, and
then to obtain the density $g_{1}\left( \alpha ^{\left( n-1\right) }\right) $
according to Eq. (29).

The numerical result of the density $g_{1}\left( \alpha ^{\left( n-1\right)
}\right) $ is shown in Fig. 3b for the range $0\leq \alpha ^{\left(
n-1\right) }\leq \pi $ with the radial distance of a zero, $r\left( \alpha
^{\left( n-1\right) }\right) $, specified in the right vertical scale. Our
results indicate that when the complex plane changes from $\tanh \eta
^{\left( n\right) }$ to $\tanh \eta ^{\left( n-1\right) }$, the distribution
density oscillates more rapidly with the peak number increasing from $2$ to $%
4$ for half cycle. The locus of the zero corresponding to the ferromagnetic
phase transition point moves from $\alpha ^{\left( n\right) }=0$ of the left 
$n$-cycle to $\alpha ^{\left( n-1\right) }=\pi $ of the $\left( n-1\right) $%
-cycle of $z_{0}=1$. For the zeros near to $\alpha ^{\left( n-1\right) }=\pi 
$, the density has the linear behavior as 
\begin{equation}
g_{1}\left( \pi +\alpha ^{\left( n-1\right) }\right) =\delta _{1}\left( 
\frac{3+2\sqrt{2}}{\pi }\right) \left| \alpha ^{\left( n-1\right) }\right| ,
\end{equation}
with 
\begin{equation}
\delta _{1}=\frac{1}{4}\left| \frac{d\alpha ^{\left( n\right) }}{d\alpha
^{\left( n-1\right) }}\right| _{\alpha ^{\left( n-1\right) }=\pi }=0.1529.
\end{equation}
This linear behavior, again, gives the logarithmic singularity of the
specific heat.

To extend the calculation of density to $\left( n-2\right) $-cycles, we may
divide the $12$ $\left( n-2\right) $-cycles shown in Fig. 4a into three
classes named as class $I$, $II$, and $III$, according to the decreasing
order from the longest to the smallest in the magnitude of the circumference
of the circles. Then, there are four members in each class, and all members
of class $I$ are the descendants of the left $n$-cycle and those beonging to
class $II$ and $III$ are from the right $n$-cycle. Similar to the case of $%
\left( n-1\right) $-cycles, we can write 
\begin{equation}
\tanh \eta ^{\left( n-2\right) }=z_{1}+r\left( \alpha ^{\left( n-2\right)
}\right) \exp (i\alpha ^{\left( n-2\right) })
\end{equation}
for the zeros of $\left( n-2\right) $-cycles, and the center coordinates, $%
z_{1}$, are chosen to be the preimages of the map of Eq. (15) for the center
coordinates of $\left( n-2\right) $-cycles, $1$, $-1$, $i$, and $-i$. Then,
we can express the densities of the $\left( n-2\right) $-cycles as 
\begin{equation}
g_{z_{1}}^{I}\left( \alpha ^{\left( n-2\right) }\right) =\frac{g_{1}\left(
\alpha ^{\left( n-1\right) }\right) }{4}\left| \frac{d\alpha ^{\left(
n-1\right) }}{d\alpha ^{\left( n-2\right) }}\right| ,
\end{equation}
and 
\begin{equation}
g_{z_{1}}^{II,III}\left( \alpha ^{\left( n-2\right) }\right) =\frac{%
g_{i}\left( \alpha ^{\left( n-1\right) }\right) }{4}\left| \frac{d\alpha
^{\left( n-1\right) }}{d\alpha ^{\left( n-2\right) }}\right| ,
\end{equation}
respectively, where the superscript, $I$, $II$, or $III$, is used to denote
the class to which a $\left( n-2\right) $-cycle belongs, and the subscript, $%
z_{1}$, is used to specify a $\left( n-2\right) $-cycle in the given class.
Since the members belonging to the same class are the same up to a global
rotation, we only need to determine the zero density of a $\left( n-2\right) 
$-cycle for each class. The cycles of $z_{1}=1$, $\sqrt{1+\sqrt{2}}e^{i\pi
/4}$, and $\sqrt{-1+\sqrt{2}}e^{i\pi /4}$, belonging to class $I$, $II$, and 
$III$ respectively, are choosen for the calculation of the density of the
respective class.

The numerical method of calculating the correspondence between $\alpha
^{\left( n-1\right) }$ and $\alpha ^{\left( n-2\right) }$ and the Jacobians, 
$\left\vert d\alpha ^{\left( n-1\right) }/d\alpha ^{\left( n-2\right)
}\right\vert $, are exactly the same as we did in the last case. The results
of $r\left( \alpha ^{\left( n-2\right) }\right) $ (right vertical scale) and 
$g\left( \alpha ^{\left( n-2\right) }\right) $ (left vertical scale) are
shown in Figs. 4b, 4c, and 4d respectively for the three cycles. These
results indicate that the rapidity of oscillation in the distribution
density of the cycles of class $I$ increases as the peak number doubles with
respect to the last case, while the peak number remains to be the same for
the cycles of class $II$ and $III$. The zero corresponding to the critical
point of ferromagnetic phase transition moves from $\alpha ^{\left(
n-1\right) }=\pi $ of the $\left( n-1\right) $-cycle of $z_{0}=1$ to $\alpha
^{\left( n-2\right) }=\pi $ of the $\left( n-2\right) $-cycle of $z_{1}=1$
of class $I$. The density of the zeros near to this locus has the linear
behavior, 
\begin{equation}
g_{z_{1}=1}^{I}\left( \pi +\alpha ^{\left( n-2\right) }\right) =\delta
_{2}\left( \frac{3+2\sqrt{2}}{\pi }\right) \left\vert \alpha ^{\left(
n-2\right) }\right\vert ,
\end{equation}
with 
\begin{eqnarray}
\delta _{2} &=&\left( \frac{1}{4}\right) ^{2}\left( \left\vert \frac{d\alpha
^{\left( n-1\right) }}{d\alpha ^{\left( n-2\right) }}\right\vert _{\alpha
^{\left( n-2\right) }=\pi }\right) \left( \left\vert \frac{d\alpha ^{\left(
n\right) }}{d\alpha ^{\left( n-1\right) }}\right\vert _{\alpha ^{\left(
n-1\right) }=\pi }\right)  \notag \\
&=&0.0365.
\end{eqnarray}

>From the densities obtained in the above, we may conjecture qualitatively
the density of the $0$-cycles with $n=4$ shown in Fig. 5 as the following:
There are $128$ members belonging to the descendants of the $\left(
n-2\right) $-cycles of class $II$ and $III$ as displayed by bold-faced
curves in Fig. 5. The $16$ members of the $128$, appearing in the outermost
of Fig. 5, are similar to a $\left( n-2\right) $-cycle of class $II$, and
the rest have a similar shape as a $\left( n-2\right) $-cycle of class $III$%
. Up to an overall reduction factor, the densities of the $16$ members have
the same oscillation pattern as that shown in Fig. 4c and the densities for
the rest members possess the same oscillation pattern as the one shown in
Fig. 4d. For the $44$ descendants of the $\left( n-2\right) $-cycles of
class $I$ displayed by gray curves in Fig. 5, the corresponding density
reduces but oscillates more rapidly with $32$ peaks in half cycle in
comparing with the one shown in Fig. 4b.

Continuing with this procedure in finding the density, we obtain $2^{n+1}$
peaks in a $\pi $\ period for each of the $0$-cycles which are the
descendants of the left $n$-cycle. On the other hand, the peak number always
maintain to be $4$ for each of the $0$-cycles belonging to the descendants
of the right $n$-cycle along with the decreasing radius as $n$ increases. \
\ \ \ \ 

\section{Julia set and infinite n limit}

The recursion relation given by Eq. (15) happens to be a rational map of
degree $4$. Then, from the work of Julia and Fatou, we may conclude that the
backward iterations defined by Eq. (16) leads towards the Julia set
associated with the map of Eq. (15).

Generally, Julia sets can be divided into two classes: Some are connected in
one piece while the others are just a cloud of points. The tendency of the
zero density with increasing $n$ shown in the above section indicates that
the Julia set here belongs to the latter. Moreover, by observing the
distribution patterns and the densities of the zeros, we may conclude that a 
$0$-cycle belonging to the descendants of the right $n$-cycle shrinks to a
point in the limit of infinite $n$, and these infinite number of points
coincide not only with the $0$-cycles generated from the left $n$-cycle but
also with the zeros obtained from the condition $\tanh \eta ^{(n-1)}=\pm i$
for infinite $n$. Thus, the same Julia set arises in DHIM as well as DDIM in
the infinite limit. In fact, the loci of the zeros of DDIM in the infinite
limit are identical to that of DHIM. \ \ \ \ \ 

The Julia set arising from the distribution of the Fisher zeros possesses
multifractal structure, and the corresponding generalized dimensions $D_{q}$
and singularity spectrum $f\left( \alpha \right) $, obtained by using
derivative method and by approximating the limiting set of the zero
distribution with that of $n=8$ shown in Fig. 6 \cite{rHL1, rHL2}, are shown
in Fig. 7.

The $\sinh \left( 2\eta \right) $ complex plane is also widely used in
studying the distribution of the zeros for simple square Ising model. The
results of simple square Ising model imply that the zeros may lie on the
unit circle, $\left\vert \sinh 2\eta ^{\left( n\right) }\right\vert =1$, for
an $n$-lattice. To obtain the distribution in the $\sinh \left( 2\eta
\right) $ complex plane, starting with the unit circle we then perform $n$%
-fold backward iterations of the map, 
\begin{eqnarray}
\sinh \left( 2\eta ^{\left( n\right) }\right) &=&\left( \frac{1}{2}\right)
\times \left[ \frac{\left( \sinh \left( 2\eta ^{\left( n-1\right) }\right)
\right) ^{4}}{1+\left( \sinh \left( 2\eta ^{\left( n-1\right) }\right)
\right) ^{2}}\right.  \notag \\
&&\left. +\frac{2\left( \sinh \left( 2\eta ^{\left( n-1\right) }\right)
\right) ^{2}}{1+\left( \sinh \left( 2\eta ^{\left( n-1\right) }\right)
\right) ^{2}}\right] ,
\end{eqnarray}%
which is an equivalent expression of Eq. (6). This is also a rational map of
degree $4$, and the Julia set associated with this map can be approximated
with $n$-fold backward iterations for a sufficiently large $n$. The
resultant distribution of $n=8$ are shown in Fig. 8. Though the distribution
pattern is different from that in the complex $\tanh \left( \eta \right) $
plane shown in Fig. 6, owing to the fact that Julia set is an invariant set
the global multifractal structure characterized by $D_{q}$ and $f\left(
\alpha \right) $ is the same as that shown in Fig. 7.

The density near the bulk transition point has a linear behavior, and the
linear behavior gives the logarithmic singularity of the specific heat. By
observing the results of Eqs. (26), (33), and (38), we may expect that the
linear behavior for the density near the bulk transition point disappears in
the limit of infinite $n$. This leads to the cusp behavior in the specific
heat at the critical point as the decoration-level goes infinite \cite{rHLL}%
. \ \ \ \ \ \ \ \ \ 

\section{summary and discussion}

\bigskip 

\label{s6} In summary, we study the distribution and the density of the
Fisher zeros for the Ising model defined on square lattices with
diamond-type bond-decorations in this paper. By carrying out the exact
renormalization map of temperature, we can express the free energy of an $n$%
-lattice as the sum of two parts: One, referred as the local part, is mainly
the contribution from an $n$-level decorated bond; and the other, referred
as the long range part, is the contribution from the interactions among $n$%
-level decorated bonds. For the local part, we obtain the zeros as a set of
scattered points. Along with the increase of the number of the scattered
points in the distribution of the zeros as the decoration-level $n$
increases, all zeros belonging to the lower decoration-levels are also
contained in the zeros distribution of the higher decoration-level. The
correponding zeros of the long-range part are the union of continuous closed
curves which also form sub-rings due to the intersections. For the
long-range part, the distribution of the zeros leads towards the Julia set
associated with the renormalization map which is a rational map of degree $4$%
. This Julia set also serves as the limiting set of the zeros obtained from
the local part of the free energy. The zero representing the critical point
of ferromagnetic phase transition is one of the zeros of the long-range part
in the physical region, and the locus of the critical point for an arbitrary
decoration-level $n$ is given. Along with the distribution pattern, we also
calculate the density of the zeros of the long-range part for the cases of $%
n=1$ and $2$. The evolution of the density of zeros in the course of
increasing $n$ indicates that the Julia set is a cloud of points. Thus, the
Julia set has the Hausdorff dimension greater than one and possesses the
multi-fractal structure.

Comparing with the continuous closed curves appearing in the zeros
distribution of the simple square Ising model, we have more complicated
distribution patterns for the bond-decorated Ising model. But, the patterns
remain to be continuous closed curves in both $\tanh \left( \eta \right) $
and $\sinh \left( 2\eta \right) $ complex planes for a finite
decoration-level $n$. This may be due to the fact that the interactions
among the Ising spins are effectively isotropic after the renormalization
map to the corresponding $0$-lattice, although the coordination numbers of
lattice sites are highly inhomogeneous for an $n$-lattice with large $n$. It
has been demonstrated on classic lattices that only when the couplings among
the nearest neighbors change from isotropic to anisotropic, the distribution
of the Fisher zeros may change from continuous curves to an area in the
plane \cite{rSK, rSC}. This leads to the conclusion that in the limit of
infinite decoration-levels the system has completely different properties:
The distribution of the zeros has multi-fractal structures, and the nature
of phase transition of the system is different from that of finite
decoration-levels \cite{rHLL}.

\section{acknowledgement}

We wish to thank Prof. F. Y. Wu for stimulating us to calculate the density
of zeros. This work was partially suppoted by the National Science Council
of ROC ( Taiwan ) under the Grand No. NSC 90-2112-M-033-002.

fig.1 A decorated bond with the decoration level (a) $n=0$, (b) $n=1$, and
(c) $n=2$. Note that the Ising spins with the Latin subscripts are referred
as inner spins.

fig. 2 (a) The distribution of the Fisher zeros of an $n$-lattice in the
complex $\tanh \left( \eta ^{\left( n\right) }\right) $ plane as the left
(gray) and the right (bold-faced) $n$-cycle, and (b) the density (left
vertical scale) and the raidus (right vertical scale) of the Fisher zeros in
the left $n$-cycle.

fig. 3 (a) The four $\left( n-1\right) $-cycles for the zero distribution of
an $n$-lattice in the complex $\tanh \left( \eta ^{\left( n-1\right)
}\right) $ plane. The cycles displayed by the bold-faced curves are the
descendants of the right $n$-circle and the gray curves are those from the
left $n$-circle. The points indicated by crosses are taken as the centers of
the $\left( n-1\right) $-cycles, and they are the preimages of the
renormalization map for the centers of two $n$-cycles. (b) The density (left
vertical scale) and the radius (right vertical scale) of the Fisher zeros in
the $\left( n-1\right) $-cycle with the center at $\left( 1,0\right) $.

fig. 4 (a) The $12$ $\left( n-2\right) $-cycles for the zero distribution of
an $n$-lattice in the complex $\tanh \left( \eta ^{\left( n-2\right)
}\right) $ plane. The points indicated by crosses are taken as the centers
of the $\left( n-2\right) $-cycles, and they are the preimages of the
renormalization map for the centers of the $\left( n-1\right) $-cycles. The $%
12$ $\left( n-2\right) $-cycles can be divided into three classes, $I$, $II$%
, and $III$, according to the decreasing order in the magitude of the
circumference of a circle. The density (left vertical scale) and the radius
(right vertical scale) of (b) the $\left( n-2\right) $-cycle, which is the
one marked $I$ in (a), with the center coordinate $(1,0)$, (c) the $\left(
n-2\right) $-cycle, which is the one marked $II$ in (a), with the center
coordinate $\sqrt{1+\sqrt{2}}e^{i\pi /4}$, and (d) the $\left( n-2\right) $%
-cycle, which is the one marked $III$ in (a), with the center coordina $%
\sqrt{-1+\sqrt{2}}e^{i\pi /4}$.

fig. 5 The zero distribution of an $n$-lattice in the complex $\tanh \left(
\eta \right) $ plane obtained from the condition of Eq. (12) for $n=4$. 

fig. 6 The zero distribution of an $n$-lattice in the complex $\tanh \left(
\eta \right) $ plane obtained from the condition of Eq. (12) for $n=8$.

fig. 7 The (a) generalized dimension $D_{q}$ and (b) singularity spectrum $%
f\left( \alpha \right) $ of the Julia set associated with the
renormalization map.

fig. 8 The zero distribution of an $n$-lattice in the complex $\sinh \left(
2\eta \right) $ plane obtained from the condition of Eq. (12) for $n=8$.


\begin{thebibliography}{99}
\bibitem{rGK} R. B. Griffiths and M. Kaufman, Phys. Rev. B \textbf{26}, 5022
(1982).

\bibitem{rSKS} N. M. Svrakic, J. Kertesz, and W. Selke, J. Phys. A \textbf{15%
}, L427 (1982).

\bibitem{rDEE} B. Derrida, J. P. Eckmann, and A. Erzan, J. Phys. A \textbf{16%
}, 893 (1983).

\bibitem{rKG} M. Kaufman and R. B. Griffiths, Phys. Rev. B \textbf{24}, 496
(1981).

\bibitem{rE} A. Erzan,\ Phys. Lett. \textbf{A93}, 237 (1983).

\bibitem{rDSI} B. Derrida, L. De. Seze, and C. Itzykson, J. Stat. Phys. 
\textbf{33}, 559 (1983).

\bibitem{rDIL} B. Derrida, C. Itzykson, and J. M. Luck, Commun. Math. Phys. 
\textbf{94}, 115 (1985).

\bibitem{rLH1} F. T. Lee and M. C. Huang, J. Stat. Phys. \textbf{75}, 1119
(1994).

\bibitem{rLH2} F. T. Lee and M. C. Huang, Chinese J. Phys. \textbf{37}, 398
(1999).

\bibitem{rMIG} A. A. Migdal, Sov. Phys. JETP \textbf{42}, 743 (1976).

\bibitem{rLPK} L. P. Kadanoff, Ann. Phys. (N.Y.) \textbf{100}, 359 (1976).

\bibitem{rBQ} A. N. Berker and S. Qstlund, J. Phys. C \textbf{12}, 4961
(1979).

\bibitem{rJKP} M. H. Jensen, L. P. Kadanoff, and I. Procaccia, Phys. Rev. A 
\textbf{36}, 1409 (1987).

\bibitem{rHL1} B. Hu and B. Lin, Phys. Rev. A \textbf{39}, 4789 (1989).

\bibitem{rHL2} B. Hu and B. Lin, Physica A \textbf{177}, 38 (1991).

\bibitem{rYL} C. Y. Yang and T. D. Lee, Phys. Rev. \textbf{87}, 404 (1952).

\bibitem{rLY} T. D. Lee and C. Y. Yang, Phys. Rev. \textbf{87}, 410 (1952).

\bibitem{rF} M. E. Fisher, \textit{Lecture Note in Theoretical Physics},
Vol. 7c, W. E. Brittin, ed. (University of Colorado Press, Boulder, 1965),
pp. 1-159.

\bibitem{rSK} W. van Saarloos and D. A Kurtze, J. phys. A: Math. Gen. 
\textbf{17}, 1301 (1984).

\bibitem{rSC} J. Stephenson and R. Couzens, Physica A \textbf{129}, 201
(1984).

\bibitem{rFEL} H. Feldmann, R. Shrock, and S. H. Tsai, Phys. Rev. E \textbf{%
57}, 1335 (1998).

\bibitem{rCHEN} C. N. Chen, C. K. Hu, and F. Y. Wu, Phys. Rev. Lett. \textbf{%
76}, 169 (1996).

\bibitem{rLW} W. T. Lu and F. Y. Wu, J. Stat. Phys. \textbf{102}, 953 (2001).

\bibitem{rHLL} M. C. Huang, Y. P. Luo, and T. M. Liaw, \textit{Ferromagnetic
phase transitions of inhomogeneous systems modeled by square Ising models
with diamond-type bond-decorations, }to be appeared in Physica A (2003).

\bibitem{rLHCL} T. M. Liaw, M. C. Huang, Y. L. Chou, and S. C. Lin, Phys.
Rev. E. \textbf{65}, 066124 (2002).
\end{thebibliography}
\end{document}